\documentclass[conference,a4paper]{IEEEtran}
\usepackage[utf8]{inputenc}
\usepackage{amsmath}
\usepackage{amsfonts}
\usepackage{amssymb}
\usepackage{graphicx}
\usepackage{xcolor}
\usepackage{soul}
\usepackage[left=1.62cm,right=1.62cm,top=1.9cm]{geometry}
\usepackage{comment}
\usepackage{caption}
\usepackage{subcaption}

\usepackage{epsfig}  
\usepackage{multirow}
\usepackage{psfrag}
\usepackage[linesnumbered,ruled]{algorithm2e}
\usepackage{float}
\usepackage{diagbox}
\usepackage{url}
\usepackage{hyperref}
\usepackage{xcolor}
\usepackage{stfloats}
\usepackage{cite}
\usepackage[normalem]{ulem}

\usepackage{tabulary}

\definecolor{boristext}{rgb}{0.22, 0.44, 0.88}
\definecolor{boriscomments}{rgb}{0.88, 0.04, 0.04}
\definecolor{boristochange}{rgb}{0.2, 0.8, 0.8}

\title{Performance Evaluation of MLO for XR Streaming: Can Wi-Fi 7 Meet the Expectations?}

\author{
\IEEEauthorblockN{Marc Carrascosa-Zamacois$^{\flat}$$^{\star}$, Lorenzo Galati-Giordano$^{\flat}$, Francesc Wilhelmi$^{\flat}$, \\Gianluca Fontanesi$^{\flat}$, Anders Jonsson$^{\star}$, Giovanni Geraci$^{\sharp}$$^{\star}$, and Boris Bellalta$^{\star}$\vspace{0.2cm}
}
\IEEEauthorblockA{$^{\flat}$\emph{Nokia Bell Labs, Stuttgart, Germany}}
\IEEEauthorblockA{$^{\star}$\emph{Universitat Pompeu Fabra, Barcelona, Spain}}
\IEEEauthorblockA{$^{\sharp}$\emph{Telef\'{o}nica Research, Barcelona, Spain}}
\thanks{M. Carrascosa and B. Bellalta were supported in part by grants MAX-R HEu-CL4-MAX-R-101070072, Wi-XR PID2021-123995NB-I00, and by MCIN/AEI under the Maria de Maeztu Units of Excellence Programme (CEX2021-001195-M).
G. Geraci was
supported in part by the Spanish Research Agency through grants PID2021-
123999OB-I00, CEX2021-001195-M, and CNS2023-145384, by the UPF-
Fractus Chair, and by the Spanish Ministry of Economic Affairs and Digital
Transformation and the European Union NextGenerationEU.}
}

\IEEEoverridecommandlockouts
\begin{document}

\bstctlcite{IEEEexample:BSTcontrol}

\maketitle


\begin{abstract}
    Extended Reality (XR) has stringent throughput and delay requirements that are hard to meet with current wireless technologies. Missing these requirements can lead to worsened picture quality, perceived lag between user input and corresponding output, and even dizziness for the end user. In this paper, we study the capability of upcoming Wi-Fi 7, and its novel support for Multi-Link Operation (MLO), to cope with these tight requirements. Our study is based on simulation results extracted from an MLO-compliant simulator that realistically reproduces VR traffic. 
    Results show that MLO can sustain VR applications. By jointly using multiple links with independent channel access procedures, MLO can reduce the overall delay, which is especially useful in the uplink, as it has more stringent requirements than the downlink, and is instrumental in delivering the expected performance.
    We show that using MLO can allow more users per network than an equivalent number of links using SLO. We also show that while maintaining the same overall bandwidth, a higher number of MLO links with narrow channels leads to lower delays than a lower number of links with wider channels.
\end{abstract}

\section{Introduction}

Extended Reality (XR) applications, which include Virtual Reality (VR) and Augmented Reality (AR), are growing in popularity as they unlock novel use cases across many domains, such as healthcare, industry, education and gaming. Most use cases are planned for indoor use, and thus Wi-Fi is expected to become the main technology to support them \cite{OUGHTON2024102766,giordano2023will}, with most headsets including high-grade Wi-Fi capabilities\footnote{\url{https://www.meta.com/help/quest/articles/headsets-and-accessories/oculus-link/connect-with-air-link/}}, and services like \emph{Steam Link}\footnote{\url{https://store.steampowered.com/app/353380/Steam_Link/}} allowing to stream games wirelessly from  computer to  headset. To deliver a good performance to the end user, XR traffic has stringent requirements in both application-level throughput, which can go over 100~Mbps, and delay, which needs to be well below 10~ms. 

Wi-Fi struggles to provide delay guarantees: Wi-Fi's operation at the MAC layer is based on distributed channel access due to its operation in the unlicensed spectrum and the inherent requirement of using Listen Before Talk (LBT). For that reason, the contention among devices associated with the sharing of the same frequency channels has a direct impact on the delay experienced by the users, which deteriorates as the number of contenders increases. Further, XR applications have stricter requirements for the uplink (UL) delay, which is harder to control by the Access Point (AP) due to the spontaneous nature of such type of traffic.

IEEE 802.11be (Wi-Fi 7) \cite{10058126, GarLopGal2021} is envisioned as an enabler for lowering network delay with Multi-Link Operation (MLO). A key feature of Wi-Fi 7, MLO allows a device to connect to multiple bands or channels through a single association and to transmit packets simultaneously over them, thus multiplying the available bandwidth by the number of radios on a device. Medium access is also independent for each radio, leading to more transmission opportunities and reduced contention as well \cite{CarGerGal22}.

Wi-Fi's capability to support XR applications has been tested in~\cite{jansen2023can}, comparing the performance of wired and wireless deployments, and showing that, while Wi-Fi can achieve similar results than a wired connection, this only happens for devices that are close to the AP and with direct line of sight. In contrast, the tests in \cite{jansen2023can} also showed that a poor Received Signal Strength Indicator (RSSI) leads to inconsistent performance and lower frame rates. In \cite{michaelides2023wi}, a setup with multiple VR users over Wi-Fi was studied, showing that scheduling the uplink transmissions (i.e., using OFDMA to improve multi-user contention) leads to worsened performance overall than just using DCF. In \cite{gonccalves2020comparative}, user experience was studied for wired and wireless VR setups, also highlighting that a direct line of sight is necessary to ensure comparable Quality of Experience (QoE) between wired and wireless setups. Different types of XR applications were studied and classified in \cite{hazarika2023towards, akyildiz2022wireless}, as well as their requirements and the possibility to cover them based on the current efforts done in 5G and Wi-Fi standardization. 

In \cite{mallik2024performance}, a performance evaluation model was proposed for edge-assisted XR applications, considering battery usage, end-to-end delay and handoff delays. In \cite{ahn2018virtual}, the authors analyzed a multi-user VR setting deployed over Wi-Fi, and concluded that Quality of Service (QoS) enforced by the standard Enhanced Distributed Channel Access (EDCA) is insufficient to support the stringent delay and packet loss requirements of such a setting. They then proposed a new architecture to improve performance by separating the downlink and uplink using the 802.11ad/ay 60 GHz band and the 802.11ax 5 GHz band, respectively.

In the particular case of Wi-Fi 7's MLO, the work in \cite{naik2021can} showed that adding a second link to traditional Wi-Fi can lead to order of magnitude gains in the 90th percentile delay for real-time applications. In \cite{carrascosa2023wi}, a dataset using real-world channel occupancy traces was used to test MLO performance, also showing a similar order of magnitude improvement over SLO in the 95th percentile delay. The performance of Wi-Fi 7 for AR applications was studied in \cite{alsakati2023performance}, concluding that MLO can serve more users than SLO with equivalent bandwidth. None of these studies, however, have considered VR traffic characteristics and performance requirements in detail.
In this paper, we focus on realistic VR streaming applications, and study Wi-Fi 7's ability to effectively meet their stringent requirements. Unlike prior work, we study the impact of VR traffic on Wi-Fi 7 networks, which we test for both Single Link Operation (SLO) and Multi-Link Operation (MLO).
Using simulation results, we showcase the relationship between different transmission parameters, namely the Modulation Coding Scheme (MCS) and channel bandwidth, and provide insights on their required configuration for achieving the desired performance for VR applications. Our analysis also studies the effect of an increasing number of users to better understand the limits of Wi-Fi 7 for VR traffic. Our main contributions are as follows:
\begin{itemize}
    \item We provide an overview of VR gaming traffic based on real traces, and analyze the associated requirements defined by the Wi-Fi Alliance (WFA).
  
    \item We conduct an extensive performance evaluation based on several simulations and discuss the feasibility of Wi-Fi 7 for supporting VR traffic. We show that the number of VR users with MLO and N links is higher than N independent SLO APs.

    \item We show that for MLO, multiple narrow links provide a better support for VR applications than few but wider links, as the channel access delay---the most limiting factor---is significantly scaled down.
\end{itemize}

\section{VR Gaming}

\subsection{VR Streaming Setup}

Streaming VR gaming applications offloads the main tasks to a server, taking the computational load of rendering the video, audio and any other necessary data away from the Head Mounted Display (HMD). The rendered video and audio is then transmitted through the internet to the HMD, which acts as a client, and reproduces the incoming video to the user, also capturing the user inputs to send back to the server. This approach is also known as split-rendering VR.

\begin{figure}[t!!]
    \centering
    \includegraphics[width=0.39\textwidth]{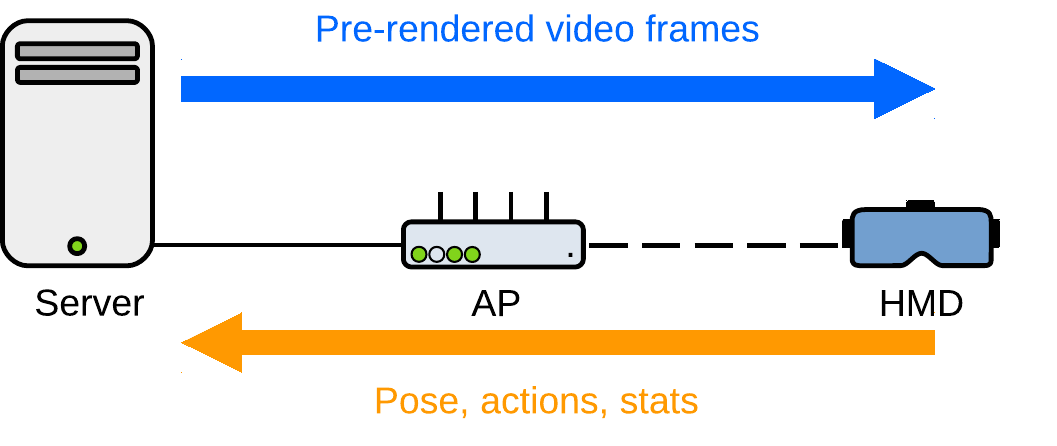}
    \caption{VR streaming components. } 
    \label{xrdiagram}
\end{figure}

In our setup, the server is connected to the AP directly via a~1 Gbps Ethernet link, and the HMD is connected to the AP wirelessly. Fig.~\ref{xrdiagram} shows the main components of VR gaming streaming in our particular setup.

\subsection{VR Traffic Distribution}

To analyze and replicate VR traffic, we use the traces from~\cite{michaelides2023wi}, which can be found in Zenodo as a dataset \cite{ds}. They were obtained using Air Light VR (ALVR), which was installed in both the server and HMD. ALVR allows the streaming of VR games over Wi-Fi, as well as gives the user control over several stream settings, such as resolution, refresh rate, codec used, and transport protocol (UDP or TCP). ALVR creates a bridge between the server and the HMD, transmitting audio, video, and tracking. Video is compressed at the server using either the H.264 or H.265 codecs. Audio is sent raw using Pulse-Code Modulation (PCM). 

\begin{figure*}[ht]
\centering
\begin{subfigure}[b]{\textwidth}
    \includegraphics[width = \textwidth]{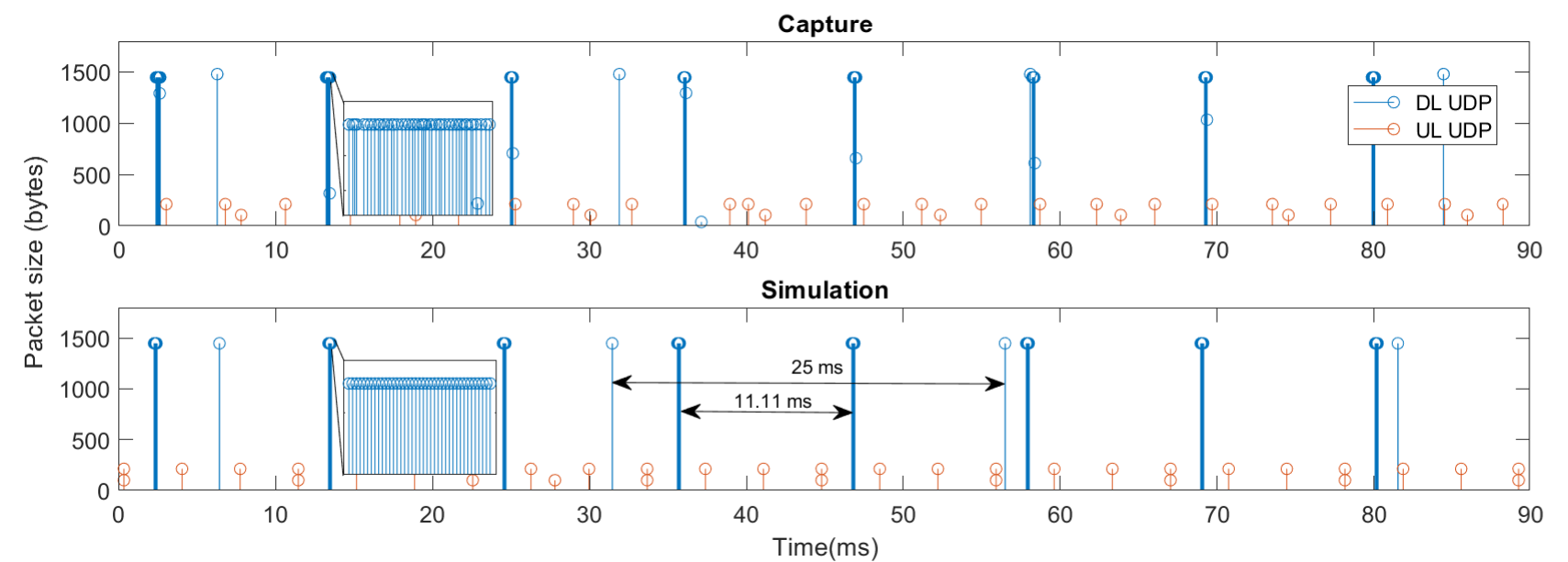}
  
\end{subfigure}

\caption{VR traffic distribution obtained from the capture (top figure) and the simulation (bottom figure).}
\label{vs}
\end{figure*}

Tests were performed at different resolutions and refresh rates, using H.264 coding and UDP for the transport protocol. Wireshark was used to capture the traffic on the server. These captures, which have been replicated in our simulations, reflect the generation patterns for VR gaming. The traffic patterns are fully described in Section~\ref{traffic}. A comparison of the captures and our simulator output is shown in Fig.~\ref{vs}. 

\subsubsection*{Downlink traffic}
For the downlink (DL) traffic there are two types of packets: video and audio. Video is transmitted in batches of packets separated as a function of the frame rate. In this case, it is 90~frames per second (FPS), which leads to an interval of 11.11~ms between batches. At a 100~Mbps application rate, each batch contains an average of 96 video packets of 1448 bytes. The audio is transmitted at a different rate (25~ms) and in batches of 4 packets. 

\subsubsection*{Uplink traffic}
In the uplink (UL), we have information about the tracking, pose, and stream statistics. They also follow the video frame rate, with 3 packets of size 106 bytes per video frame, and a single one of 212 bytes, which we believe accounts for the pose and stats, respectively.

\subsection{Throughput and Delay Requirements}

VR content has strict throughput and delay requirements. It requires real-time rendering, meaning that a large amount of data needs to be transmitted constantly and consistently, and buffering is not possible. VR frame rates are high, ranging from 72 FPS to 120 FPS. This frame rate sets the pace at which traffic is generated, and so the higher the quality of the stream, the more frequent the transmissions. Video quality is also affected by its bitrate, which can range from 40 to 200 Mbps. The delay is particularly important as well, not only to deliver a good video experience, but to avoid that the user suffers dizziness.  Finally, the rendered video in the downlink changes based on the inputs of the user, which are delivered by the uplink, thus it is important to protect the uplink so that the downlink is displaying the correct output in a timely manner. In this work, we will look at the delay thresholds set by the Wi-Fi Alliance for VR gaming~\cite{wifialliance}, which we summarize in Table~\ref{teVar2}.

\setlength\tabcolsep{7 pt}
	\begin{table}[t]\centering
	\caption{Delay requirements defined by Wi-Fi Alliance for VR gaming.}
	\begin{small}
 \renewcommand{\arraystretch}{1.15}
    \begin{tabular}{|c|c|c|}
  		\hline 
		\multirow{ 3}{*}{\textbf{ Type of Traffic Stream}}   &   \textbf{Required} & \textbf{Maximum}\\
   & \textbf{Reliability}& \textbf{Recommended}\\
   & \textbf{(Percentile)}&\textbf{Delay (ms)}\\\hline
        \multirow{3}{*}{Video frames (DL)} & 75th & $5$ \\\cline{2-3}
         & 95th &$10$\\\cline{2-3}
         & 99.9th & $50$\\\hline
         Pose, IMU & 90th &$2$\\\cline{2-3}
         Controller inputs (UL)& 99.9th&$10$\\\hline
	\end{tabular}
  \label{teVar2}

	\end{small}
 
	\end{table}

\subsection{Challenges}

\section{System Model}

\subsection{VR Traffic Characterization}\label{traffic}

VR traffic is periodic, defined mainly by its total traffic load and the frame rate used by the VR application. As feedback from the HMD is continuously transmitted to the server, a constant video bitrate is generated without exception.  We match these main characteristics in our simulation: the frame rate $\phi$ sets the inter-arrival time $\Delta$ of the video data, with downlink packet batches separated by $\Delta = \frac{1}{\phi}$ secs. The video bitrate $\rho$ sets the size of the downlink video batches (in packets per batch), which corresponds to $N_{\text{batch}} = \Delta\frac{\rho}{\text{L}}$, where $\text{L}$ is the video packet size.

\subsection{Channel Access}
\begin{figure}
    \centering
    \includegraphics[width=0.45\textwidth]{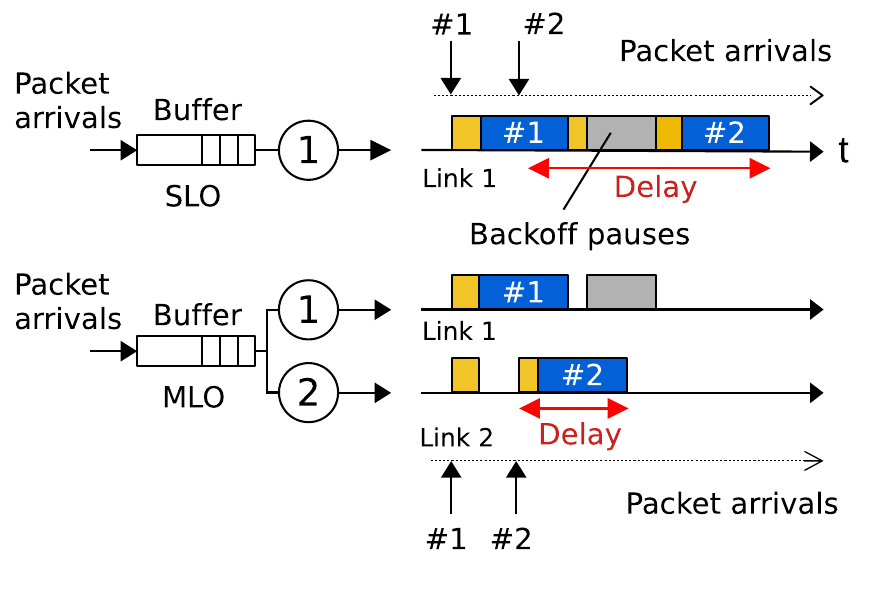}
    \caption{Channel access modes.}
    \label{xrdiagram2}
\end{figure}
We consider two main modes of operation (represented in Fig. \ref{xrdiagram2}):
\begin{itemize}
    \item \textbf{Single Link Operation (SLO):} Current Wi-Fi operation, used as our baseline. APs and STAs connect through a single link and then perform backoff to access it. Packets are transmitted sequentially.
    \item \textbf{Multi-Link Operation STR (MLO):} MLO Simultaneous Transmit and Receive (STR)\footnote{For the remainder of the paper, we use MLO to refer to MLO STR operation.} allows APs and STAs to connect through multiple channels at the same time. Each link uses an independent backoff timer, thus packets can be transmitted opportunistically through both links. Fig.~\ref{xrdiagram2} shows packet~\#2 arrives at the buffer once packet~\#1 is in the middle of being transmitted through link~1. Backoff is then performed in the second link, and the packet is transmitted at the same time as packet~1, reducing the delay in comparison to the sequential transmission in SLO.
\end{itemize}

\subsection{Scenario}

We consider a single Basic Service Set (BSS) Wi-Fi network. It consists of one AP and $K$ VR stations. The VR server has a direct cabled connection to the AP. We consider Wi-Fi 7 modulation and coding schemes (up to 4096-QAM), and path loss at 5 GHz band is modeled considering the 802.11ax residential scenarios~\cite{path}. Simulation parameters can be found in Table \ref{teVar3}.

\setlength\tabcolsep{7 pt}
	\begin{table}[t]\centering
	\caption{Simulation parameters.}
	\begin{small}
 \renewcommand{\arraystretch}{1.15}
    \begin{tabular}{|c|c|}
  		\hline 
		\textbf{ Name}   &   \textbf{Value}\\\hline
 
        Channel Bandwidth & 20, 40, 80, 160, 320 MHz \\\hline
         Transmission Power & 23 dBm \\\hline
         Clear Channel Assessment & -82 dBm \\\hline
         Spatial streams & 2 \\\hline
         PER & 10\% \\\hline
         Max. Packet Aggregation & 1024 A-MPDU \\\hline
         Buffer size & 5000 packets \\\hline
         Iterations & 100 seeds\\\hline
         Simulation time & 10 seconds \\\hline
    
	\end{tabular}
  \label{teVar3}

	\end{small}
 
	\end{table}

\section{Performance Evaluation}

\subsection{Required MCS and Channel Bandwidth}
\begin{figure*}[ht]
\centering
\begin{subfigure}[b]{0.45\textwidth}
    \includegraphics[width = \textwidth]{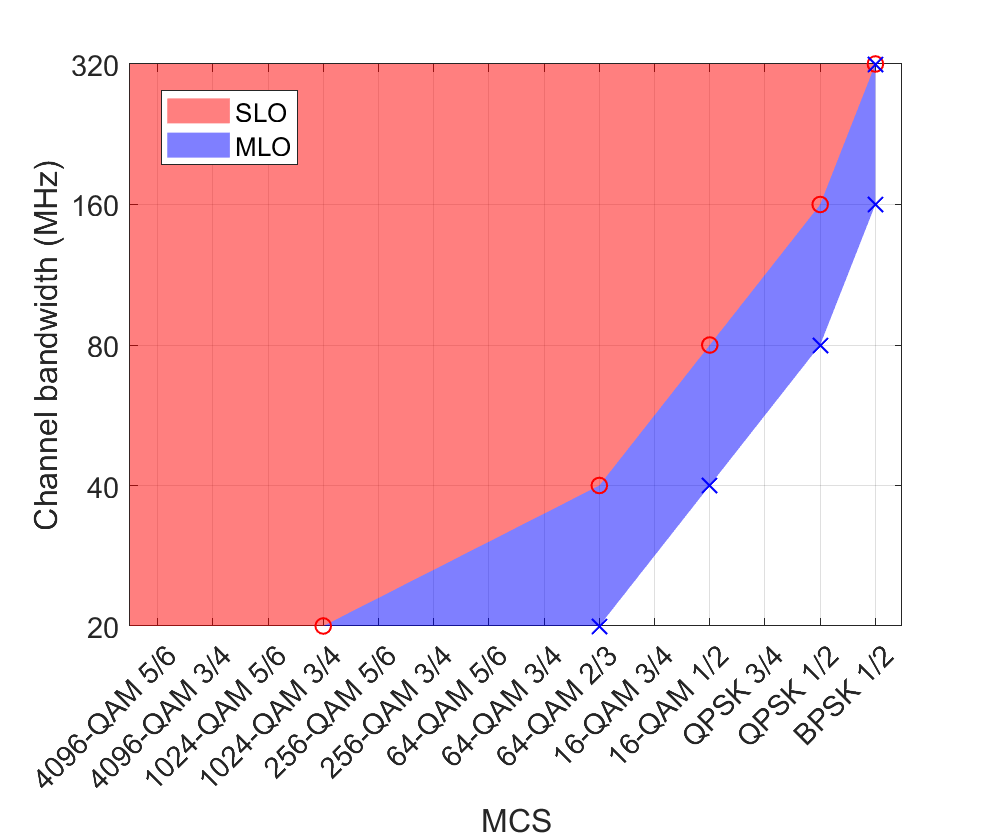}
    \caption{Downlink (75\%-tile of 5 ms)}
    \label{dl}
\end{subfigure}
\begin{subfigure}[b]{0.45\textwidth}
    \includegraphics[width = \textwidth]{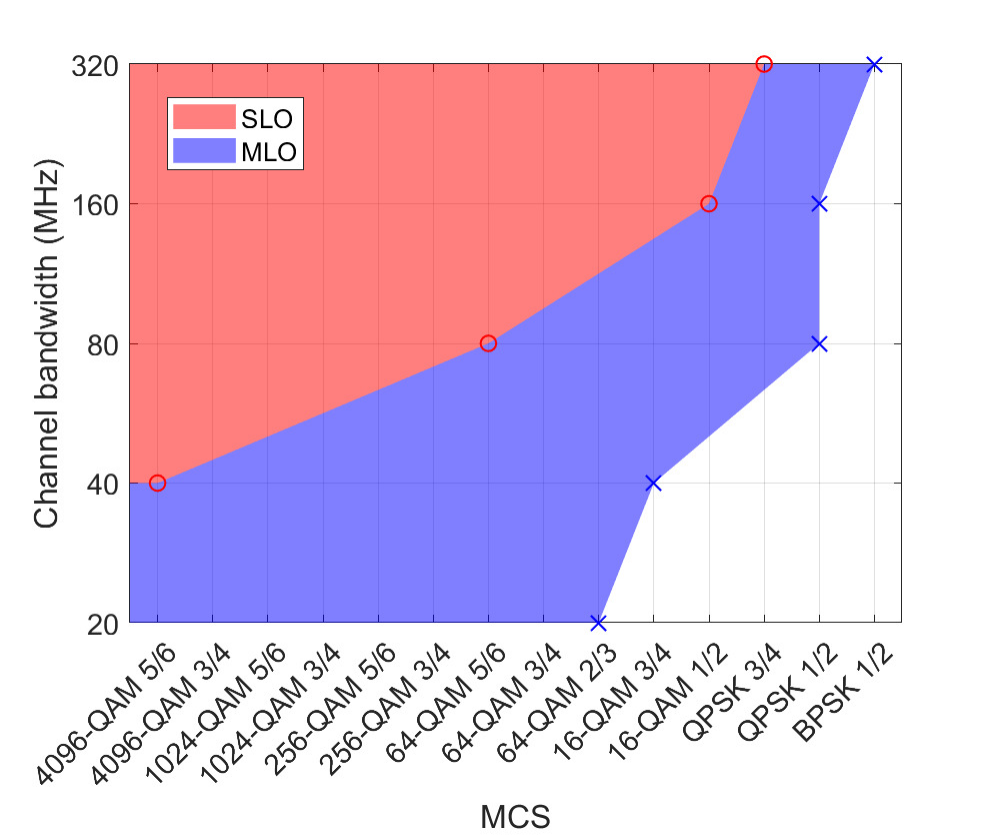}
    \caption{Uplink (90\%-tile of 2 ms)}
    \label{ul}
\end{subfigure}

\caption{Minimum MCS to accomplish Wi-Fi Alliance thresholds for different channel widths.} 
\label{mcs}
\end{figure*}

We start our study by verifying the minimum combination of MCS and channel bandwidth required for delivering a good experience to the end user according to Wi-Fi Alliance specification for XR gaming \cite{wifialliance}, defined in Table~\ref{teVar2}. 
To do so, we compare the ability of Wi-Fi 7 MLO with respect to previous generation SLO only.  
We consider a single AP and STA operating at different channel bandwidths, from 20 MHz to 320 MHz. For each bandwidth, different MCS values are evaluated to find the combinations that meet a good user experience. 

Fig.~\ref{mcs} compares the results obtained with latest Wi-Fi 7 MLO and SLO-only devices for both downlink (Fig. \ref{dl}) and uplink (Fig.~\ref{ul}). 
For SLO, the uplink is far more restrictive than the downlink and a minimum bandwidth of 40 MHz associated to an extremely high MCS, i.e., 4K QAM, is required. In the downlink, VR requirements can be met even with 20 MHz and 1024-QAM with coding rate 3/4. For all other bandwidth configurations, the UL in SLO requires much higher MCS values than the downlink, and even with 320 MHz, the lowest MCS cannot achieve the 2~ms requirement at 90th percentile. 
On the contrary, for MLO the disparity between uplink and downlink is much lower, generally requiring similar MCS values, which are also lower than the ones required by SLO.

\textbf{\textit{Takeaway:}} The UL requirements are harder to meet than the DL despite having a much lower traffic load. 
MLO offers a clear advantage over SLO even in situations where the total used bandwidth is comparable with SLO (e.g., SLO using 160 MHz and MLO using two links of 80 MHz), as MLO implicitly relieves the UL/DL self-contention by taking advantage of transmitting over multiple links. 
In addition, by demanding lower MCS with the same total bandwidth, MLO provides a better flexibility with respect to SLO for VR streaming in more challenging propagation conditions, such as being far away from the AP or not in direct line of sight.  

\subsection{Increasing Number of VR Users with MLO}

We now study the capacity of MLO to serve a certain number of VR streams and compare it to legacy SLO. 
We set a single AP transmitting multiple VR streams of 100 Mbps and 90 FPS to end users. All STAs have the same MCS of 1024-QAM 5/6 over 80 MHz channels.

\begin{figure*}[ht]
\centering
\begin{subfigure}[b]{0.455\textwidth}
    \includegraphics[width = \textwidth]{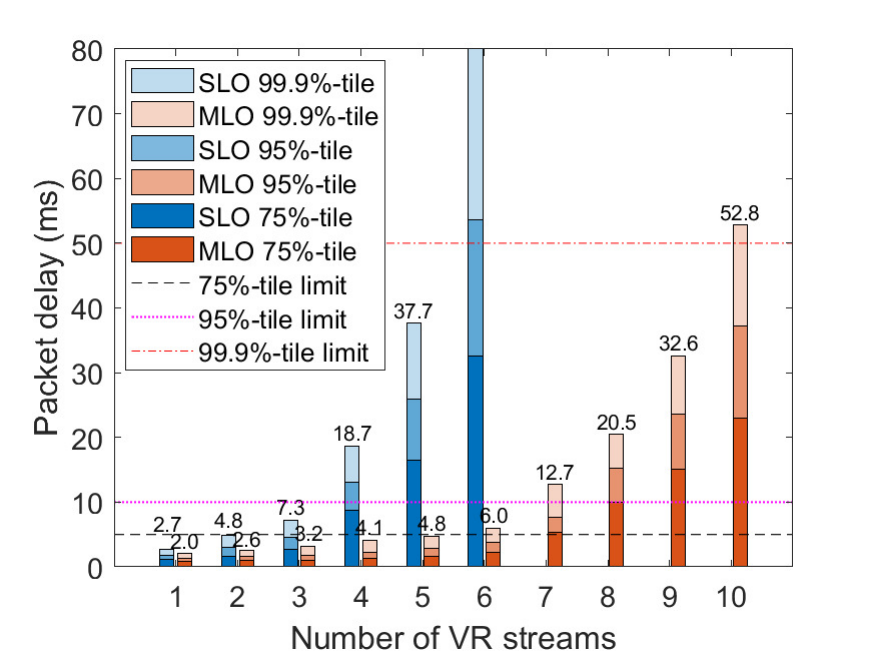}
   \caption{Downlink packet delay}
    \label{dlpackdelay100}
\end{subfigure}
\begin{subfigure}[b]{0.455\textwidth}
    \includegraphics[width = \textwidth]{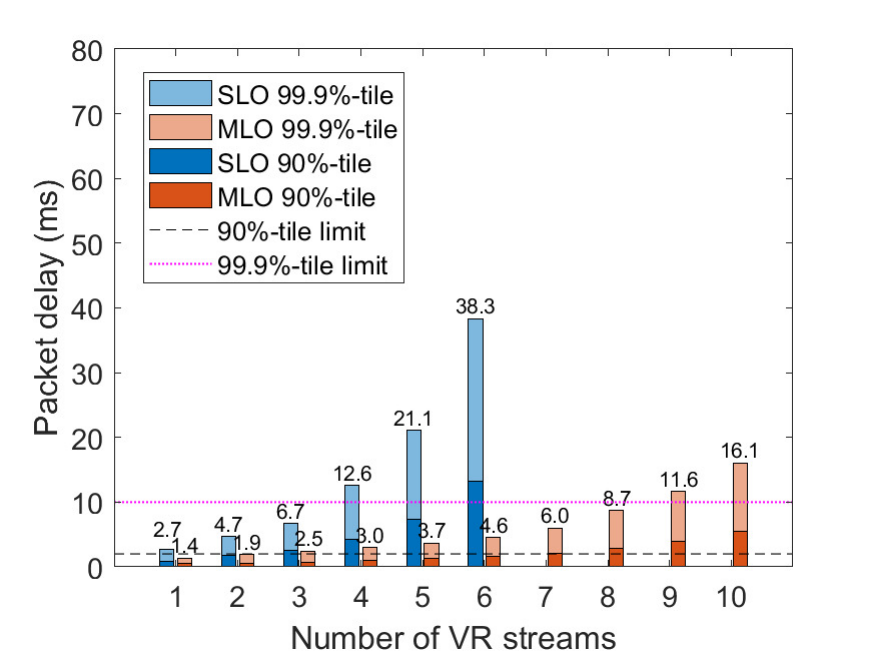}
   \caption{Uplink packet delay}
    \label{ulpackdelay100u}
\end{subfigure}
\caption{Packet delay as number of users increases. The straight lines indicate the respective Wi-Fi Alliance requirements for DL and UL.}
\label{xrstreamspd}
\end{figure*}

Fig.~\ref{dlpackdelay100} shows the 75th, 95th and 99.9th percentiles of the packet delay suffered by DL traffic, for both SLO and MLO, and for an increasing number of VR streams. The dashed lines highlight the DL delay thresholds for each percentile (as defined in Table \ref{teVar2}). We can observe that SLO's delay increases much faster than MLO's, allowing three streams
before exceeding the 75th percentile threshold of 5~ms (black dashed line). MLO comfortably allows up to six streams, and offers delay improvements at lower loads (e.g., for three streams, SLO has a 99.9\% delay of 7.3 ms, while MLO has a delay of 6 ms for six streams). Generally, the number of extra streams that can be added with MLO is directly proportional to the extra number of links enabled. However, in all cases, MLO guarantees a lower delay in both DL and UL compared to SLO while supporting twice the number of VR streams, showing that the opportunistic nature of MLO offers a slight improvement to network delay.

Similar to the DL, Fig.~\ref{ulpackdelay100u} shows the 90th and 99.9th percentiles for the packet delay at the UL. It can be observed that for SLO, the UL does not allow more than two users, exceeding the 2 ms threshold for the third user, once again showing that the UL limits SLO connections. In contrast, MLO can sustain six users, the same number as in the downlink. Additionally, the 99.9th percentile delay for six MLO users is 4.6 ms, which is lower than the 4.7 ms achieved by SLO with two users. This indicates that even if we had two SLO networks to match the bandwidth, SLO would only support four VR users, while MLO allows for an extra 50\%.

\textbf{\textit{Takeaway:}} The MLO advantage over SLO is a consequence of the increase in the number of independent channel access instances over the available links, rather than the increased bandwidth, allowing MLO to sustain more VR users than SLO deployments in non-overlapping bands using the same total bandwidth.


\subsection{Configuring MLO for VR applications}

We now attempt to further increase the number of VR users in the network by reducing contention, distributing the same bandwidth over a different number of links, thus increasing the opportunities for streams to be transmitted in parallel. MLO configurations of two, four and eight links of 80~MHz,  40~MHz and 20~MHz are used respectively, maintaining overall bandwidth used, but spreading it over an increasing number of links.

\begin{figure}[ht]
\centering
\begin{subfigure}[b]{0.46\textwidth}
    \includegraphics[width = \textwidth]{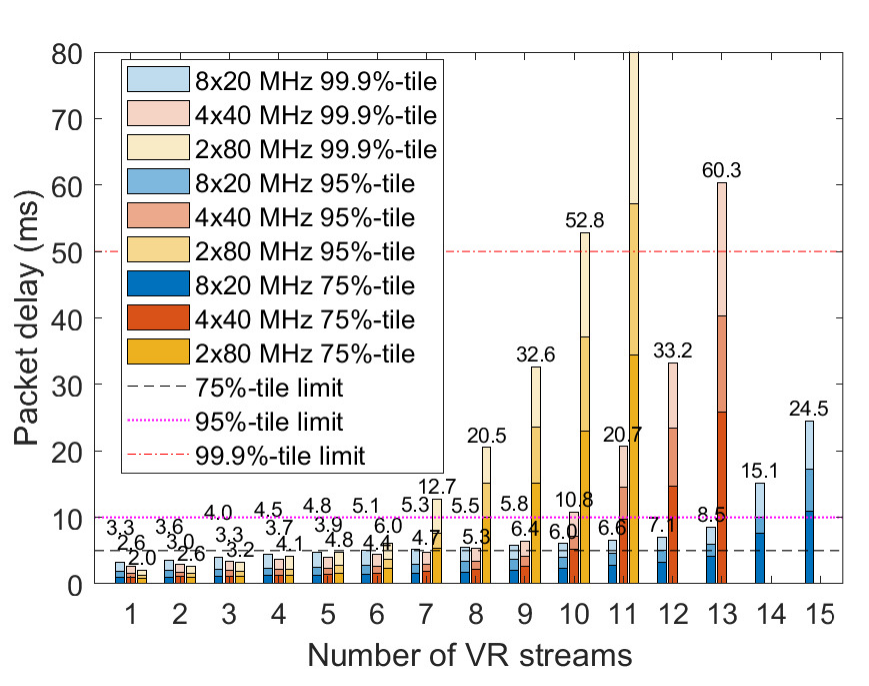}
    \caption{Downlink}
    \label{dlb}
\end{subfigure}
\begin{subfigure}[b]{0.46\textwidth}
    \includegraphics[width = \textwidth]{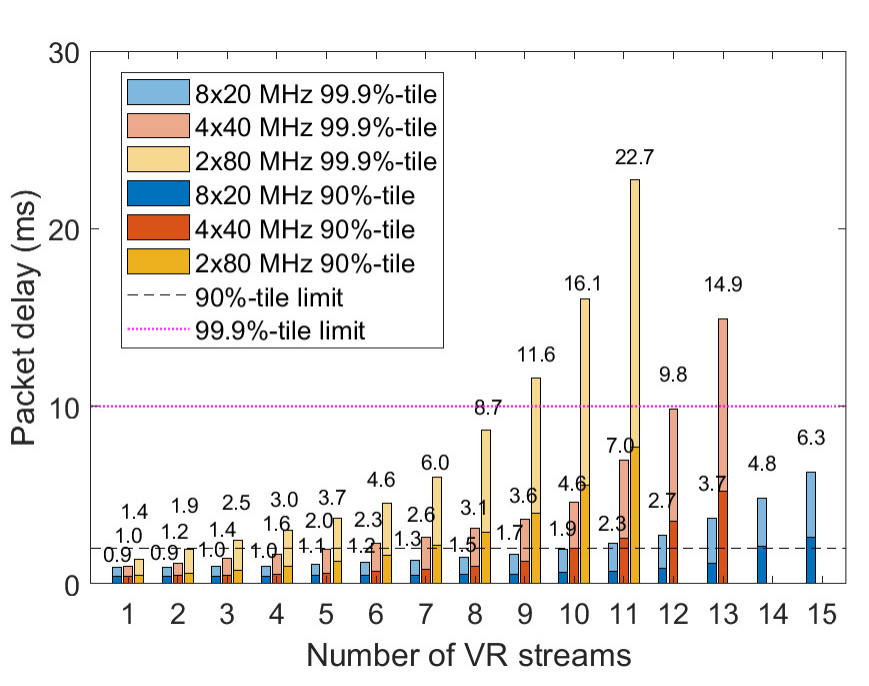}
   \caption{Uplink}
    \label{ulb}
\end{subfigure}
\caption{Packet delay for different configurations of links and bandwidth.  }
\label{f222}
\end{figure}

Fig.~\ref{dlb} shows the downlink packet delay for all configurations. We can observe that adding more links, even if each has lower bandwidth, can increase the number of VR users supported. With two links we get to serve up to six users, with four links we can serve up to nine users, and eight links allow supporting up to thirteen users. Note that if we focus on the cases where all configurations meet the requirements, for up to three users, two links of 80 MHz result in lower delays overall. Then from four to eight users, four links of 40 MHz is the best option. Beyond nine users, it is better to use eight links. 

Fig.~\ref{ulb} shows the uplink packet delay for all three configurations, in which we can observe that increasing the number of links improves the delay for all cases. 
These results show there is a clear trade-off between the bandwidth used per channel and the links-per-user ratio. When we have few users but many links, as the downlink arrives in batches, most links end up being idle while a subset are used for transmitting all the data. In these cases, a higher bandwidth allows for higher data capacity and lower transmission times. Once the number of users increases, we have a higher chance of transmitting simultaneously on all the links, and having fewer links results in increased waiting times in the queue. If we have four users and two links, we can only support up to two users simultaneously. Once there are packets from more users than links, some packets must wait for the ongoing transmission to finish before being transmitted, thus leading to increased delay.

The uplink behaves differently due to two reasons: the first is the lower traffic load required per STA, and the second is the timing between packets. As uplink packets do not arrive in batches, the buffer does not fill as quickly as the downlink, resulting in minimal aggregation. Transmissions are always short, thus not benefiting from higher capacity, and having more links allows all packets to be sent as soon as possible.

\textbf{\textit{Takeaway:}} There is a trade-off in the DL between channel access opportunities and transmission time (more independent links vs. more bandwidth per link). This is also affected by the number of VR users in the network. 
Under the assumption of using the same total bandwidth, the number of configured links should be large enough to guarantee that the delay-sensitive UL transmissions are not blocked by channel access contentions and, at the same time, maintain a sufficient bandwidth to support the high DL throughput demand of VR applications.

\section{Conclusions}

In this paper, we modeled real VR traffic traces to test Wi-Fi MLO capabilities to support the stringent requirements of VR traffic in terms of MCS-bandwidth pairs and number of links. 
We showed that MLO can support VR applications with lower bandwidth and lower MCS than SLO, providing more robustness over a wider range of propagation conditions. 
We also showed that MLO offers lower delays due to increasing the number of independent channel access instances and that using an equivalent number of links, MLO allows an extra 50\% of users per network over SLO.
In order to accommodate a higher number of VR users, a proper configuration of links and channel bandwidth is required. Spreading the same bandwidth over more links can allow for more users to contend in the network without exceeding delay requirements, but for a lower user count, having an excess of links may not result in any gains.

In future work, we intend to further dive into ways to utilize MLO to further increase delay gains, as well as study coexistence between MLO and SLO devices. Wi-Fi 8 \cite{giordano2023will} will also bring even further improvements that could be used to drive this type of content, such as Multi-AP coordination, allowing to reduce contention between VR users associated to different APs.

\bibliographystyle{IEEEtran}
\bibliography{main.bbl}

\end{document}